\documentclass[onecolumn,nofootinbib,nobibnotes,notitlepage]{revtex4-1}

\usepackage{hyperref}
\usepackage{enumerate}
\usepackage{amsmath,amssymb,amsfonts}
\usepackage{cleveref}
\usepackage{xcolor}
\usepackage[english]{babel} 
\usepackage[utf8]{inputenc}
\usepackage{mathrsfs}
\usepackage{graphicx}
\usepackage{array,multirow}
\usepackage{adjustbox}
\usepackage{dsfont}
\usepackage{textcomp}

\crefname{equation}{Eq.}{Eqs.}
\crefname{figure}{Fig.}{Figs.}
\crefname{section}{Sec.}{Secs.}
\crefname{table}{Tab.}{Tabs.}
\crefname{appendix}{Appx.}{Appx.}
\Crefname{equation}{Equation}{Equations}
\Crefname{figure}{Figure}{Figures}
\Crefname{section}{Section}{Sections}
\Crefname{table}{Table}{Tables}
\Crefname{appendix}{Appendix}{Appendices}

\begin{document}
\title{Scatterometry Measurements with Scattered Light Imaging Enable New Insights into the Nerve Fiber Architecture of the Brain}

\author{Miriam Menzel$^{1,\ast}$}
\author{Marouan Ritzkowski$^1$}
\author{Jan Andr\'e Reuter$^1$}
\author{David Gr\"a{\ss}el$^1$}
\author{Katrin Amunts$^{1,2}$}
\author{Markus Axer$^1$}

\affiliation{$^{1}$Institute of Neuroscience and Medicine (INM-1), Forschungszentrum Jülich GmbH, Jülich, Germany\\
	$^{2}$Cécile and Oskar Vogt Institute of Brain Research, University Hospital Düsseldorf, University of Düsseldorf, Düsseldorf, Germany}

\email{m.menzel@fz-juelich.de}


\begin{abstract}
The correct reconstruction of individual (crossing) nerve fibers is a prerequisite when
constructing a detailed network model of the brain. The recently developed technique
Scattered Light Imaging (SLI) allows the reconstruction of crossing nerve fiber pathways
in whole brain tissue samples with micrometer resolution: the individual fiber orientations are determined by illuminating unstained histological brain sections from different directions, measuring the transmitted scattered light under normal incidence, and studying the light intensity profiles of each pixel in the resulting image series. So far, SLI measurements were performed with a fixed polar angle of illumination and a small
number of illumination directions, providing only an estimate of the nerve fiber directions and limited information about the underlying tissue structure. Here, we use a display with individually controllable light-emitting diodes to measure the full distribution of scattered light behind the sample (scattering pattern) for each image pixel at once, enabling scatterometry measurements of whole brain tissue samples. We compare our results to coherent Fourier scatterometry (raster-scanning the sample with a non-focused laser beam) and previous SLI measurements with fixed polar angle of illumination, using sections from a vervet monkey brain and human optic tracts. Finally, we present SLI scatterometry measurements of a human brain section with 3\,\textmu m in-plane resolution, demonstrating that the technique is a powerful approach to gain new insights into the nerve fiber architecture of the human brain.
\end{abstract}

\maketitle


\section{Introduction}

Disentangling the highly complex and densely grown nerve fiber network in the brain is key to understanding its function and to developing treatments for neurodegenerative diseases. Especially the detailed reconstruction of crossing, long-range nerve fiber pathways in densely packed white matter regions poses a major challenge for many neuroimaging techniques. Diffusion magnetic resonance imaging allows to measure the spatial orientations of crossing nerve fibers, but only with resolutions down to a few hundred micrometers in post-mortem human brains \cite{calabrese2018,roebroeck2018}, which is not sufficient to resolve individual nerve fibers with diameters in the order of 1\,\textmu m \cite{liewald2014}. 
The microscopy technique \textit{3D Polarized Light Imaging (3D-PLI)}, on the other hand, determines the three-dimensional course of nerve fiber pathways in whole histological brain sections with in-plane resolutions of 1.3 \textmu m, but yields only a single fiber orientation vector for each measured tissue voxel \cite{MAxer2011_1,MAxer2011_2,dohmen2015,menzel2015}, leaving uncertainties if the brain section (with a thickness between 20\,\textmu m and 100\,\textmu m) contains several crossing nerve fibers.

Recent studies \cite{menzel2020,menzel2020-BOEx} have shown that nerve fiber crossings can be visualized with scattered light: When shining light in the optical regime through unstained, histological brain sections and studying the spatial distribution of scattered light behind the sample (\textit{scattering pattern}), we obtain valuable information about the tissue substructure of the illuminated region, such as the individual directions of crossing nerve fibers.

One possibility to measure these scattering patterns is to illuminate the sample by a non-focused laser beam, place the camera in the back-focal plane of the lens, and measure the Fourier transform of the image plane (\textit{coherent Fourier scatterometry}, \cite{menzel2020-BOEx}). However, the technique demands raster-scanning of the brain section, and the minimum diameter\footnote{The diameter of the laser beam is determined by the diameter of the pinhole. To avoid diffraction artifacts and ensure that the sample is illuminated by an approximately plane wave, the pinhole needs to be much larger than the wavelength of light. Scattering patterns measured with pinholes smaller than 100\,\textmu m were shown to provide not enough detail to reliably determine individual nerve fiber orientations.} of the laser beam ($>$ 100 \textmu m) limits the spatial resolution. 

The recently developed neuroimaging technique \textit{Scattered Light Imaging (SLI)} uses a reverse setup: Instead of measuring the scattering patterns for each brain region separately, the whole brain section is illuminated from different angles and the transmitted light is measured under normal incidence \cite{menzel2021}. In contrast to coherent Fourier scatterometry, SLI can be performed with commercial LED light sources and reconstructs crossing nerve fiber directions\footnote{Although SLI can determine the in-plane directions of multiple crossing nerve fibers, it cannot discern whether one fiber is above or below another one.} for whole brain tissue samples with micrometer resolution. So far, SLI measurements have been performed with a fixed polar angle of illumination of about $\theta \approx 47^{\circ}$ and azimuthal steps of $\Delta\phi=15^{\circ}$ (\textit{angular SLI measurement}, \cite{menzel2021}). Each pixel in the resulting image series is associated with a light intensity profile that shows peaks at different positions, revealing the directions of crossing nerve fibers. The software \textit{SLIX (Scattered Light Imaging ToolboX}, \cite{slix}) enables the generation of human-readable parameter maps, showing e.\,g.\ the individual orientations of several crossing nerve fibers. 
While 3D-PLI yields a single fiber orientation for each measured tissue voxel, angular SLI was shown to reveal the in-plane directions of up to three crossing nerve fiber bundles with $\pm 2.5^{\circ}$ accuracy (for $\Delta\phi=15^{\circ}$, see \cite{menzel2021}), providing valuable additional information for fiber tractography algorithms. Also in other regions with low 3D-PLI signal (e.\,g.\ out-of-plane fibers), SLI can serve as validation in order to obtain more reliable fiber directions.
However, the light intensity profiles obtained from angular SLI measurements contain much less information than the complete scattering patterns (especially for non-radially symmetric scattering patterns), and the fiber orientations cannot be reliably determined at the borders of the image due to asymmetric illumination.

Here, we present \textit{SLI scatterometry}, which allows measurements of full scattering patterns for each image pixel at once: Making use of an LED display with individually controllable LEDs (instead of the masked light source used in previous angular SLI measurements), we realize scatterometry measurements for whole brain tissue samples, providing detailed information about the tissue substructure and enabling a more exact determination of the individual (crossing) nerve fiber directions. 
We show that the measured scattering patterns are comparable to those obtained from previous measurements with coherent Fourier scatterometry \cite{menzel2020-BOEx}, agree with predictions from simulation studies \cite{menzel2020}, and are compatible with fiber directions derived from angular SLI measurements \cite{menzel2021}, see Sec.\ \ref{sec:comparison}.
We demonstrate that, in contrast to the line profiles obtained from angular SLI, the scattering patterns provide full structural information of the tissue (including information about the out-of-plane fiber angle) and yield reliable fiber orientations also at image borders because the center position of the patterns can be easily determined.
Finally, we present SLI scatterometry measurements of a human brain section with 3 \textmu m in-plane resolution (Sec.\ \ref{sec:human-section}), and demonstrate that our technique enables new insights into the nerve fiber architecture of human brain tissue structures.


\section{Materials and Methods}

\subsection{Preparation of Brain Tissue}
\label{sec:methods-preparation}

The studies were performed on various post-mortem brain sections.
For the comparison in Sec.\ \ref{sec:comparison}, the exact same tissue samples were used as in \cite{menzel2020-BOEx} and \cite{menzel2021}: Two 60\,\textmu m thin, coronal sections of a vervet monkey brain (sections no.\ 493 and 512), and three 30\,\textmu m thin sections of human optic tracts that were manually placed on top of each other to obtain a model of three crossing nerve fiber bundles (sections no.\ 32/33), cf.\ Tab.\ F1 in \cite{menzel2021}. 
The vervet monkey brain was obtained from a healthy adult male (2.4 years old) in accordance with the Wake Forest Institutional Animal Care and Use Committeee (IACUC \#A11-219). Euthanasia procedures conformed to the AVMA Guidelines for the Euthanasia of Animals. The optic tracts were extracted from the optic chiasm of a human brain (female body donor, 74 years old, without known neurological/psychiatric disorders) and cut along the fiber tracts of the visual pathway. 
The coronal section of the human brain hemisphere shown in Sec.\ \ref{sec:human-section} is 50\,\textmu m thin and was obtained from a female body donor (89 years old, without known neurological/psychiatric disorders).
The human brains were provided by the Netherlands Brain Bank, in the Netherlands Institute for Neuroscience, Amsterdam. A written informed consent of the subjects is available.

Within 24 hours after death, the brains were removed from the skull and fixed in a buffered solution of 4\,\% formaldehyde in which they remained for several weeks. Subsequently, they were cryoprotected in a solution of 20\,\% glycerin, deeply frozen, cut with a large-scale cryostat microtome (\textit{Polycut CM 3500, Leica Microsystems}, Germany), mounted on glass slides, and stored at -80\,$^{\circ}$C. Prior to imaging, the brain sections were thawed, embedded in 20\,\% glycerin solution, and cover-slipped.
\\


\subsection{Measurement Setup}
\label{sec:methods-setup}

\begin{figure}[h!]
\begin{center}
\includegraphics[width=0.5\textwidth]{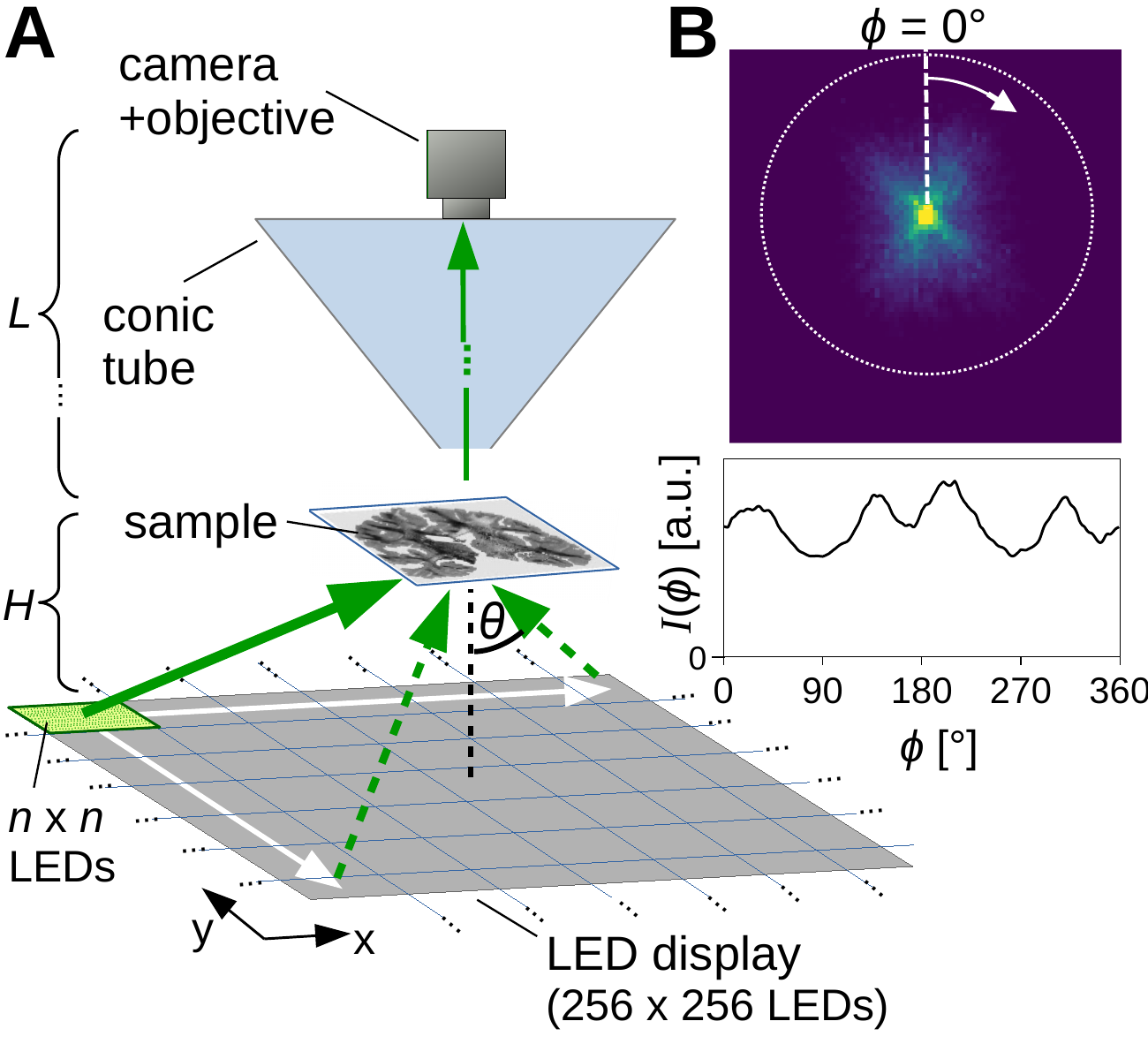}
\end{center}
\caption{\textbf{(A)} Setup for SLI scatterometry measurement. \textbf{(B)} Example of measured scattering pattern (obtained from an SLI scatterometry measurement with $2\times2$ illuminated LEDs and $81 \times 81$ kernels, shown in Fig.\ \ref{fig:chiasm}C(ii)) with corresponding azimuthal line profile (integrated from the center to the outer border, evaluated in steps of $\Delta\phi=1^{\circ}$).}
\label{fig:setup}
\end{figure}

Figure \ref{fig:setup}A shows the setup for the SLI scatterometry measurements. 
It consists of an LED display (\textit{INFiLED s1.8 LE Indoor LED Cabinet}) comprising $256 \times 256$ individually controllable RGB-LEDs with a pixel pitch of 1.8\,mm and a sustained brightness of 1000 cd/m$^2$, a specimen stage, and a camera (\textit{BASLER acA5472-17uc}) with $5472 \times 3648$ pixels and $13.1 \times 8.8$\,mm$^2$ sensor size. 
The distance between sample and camera objective was chosen to be about $L=40$\,cm, the distance between light source and sample $H=13$\,cm. The width of the LED display and the distance between light source and sample determine the maximum possible angle of illumination: $\theta = \arctan\left(\frac{256\,\cdot\,1.8\,\text{mm} / 2}{13\,\text{cm}}\right) \approx 60.6^{\circ}$.
The measurements were performed using an objective lens (\textit{Rodenstock Apo-Rodagon-D120}) with a focal length of 120\,mm and a full working distance of 24.3\,cm, yielding an object-space resolution of 3.0\,\textmu m/px and a field of view of 16.1 $\times$ 11.0\,mm$^2$. 
To avoid detection of ambient light and suppress internal reflexes, the light path between sample and camera objective was sheltered by a light-absorbing conic tube (cf.\ Fig.\ \ref{fig:setup}A).
In addition, a light-absorbing mask was placed on top of the specimen stage to cover everything outside the field of view and suppress reflexes in the camera objective (not shown in the figure).
During a measurement, a square of $n \times n$ illuminated RGB-LEDs (white light) was moved over the LED display ($m \times m$ kernels) and an image was taken for every position of the square, yielding a series of $m \times m$ images. Depending on the size of the illuminated square, the measurements were performed for different times of illumination (from 100\,ms to 10\,s) and different camera gain factors (from 1 to 27).
To reduce noise, up to four shots were recorded for the same position of illuminated LEDs and averaged.
\\

\subsection{Generation of Scattering Patterns and SLI Profiles}
\label{sec:methods-scatteringpatterns}

For every image pixel in the resulting series of $m \times m$ images, a scattering pattern with $m \times m$ pixels can be generated (cf.\ Fig.\ \ref{fig:setup}B, top): The pixel value in the upper left corner of the scattering pattern, for example, shows the intensity value of the respective image pixel in the first recorded image (obtained when illuminating the sample with $n \times n$ LEDs in the upper left corner of the display).
In this way, the scattering pattern shows the distribution of scattered light for the respective image pixel. 
When comparing the scattering patterns to measurements with coherent Fourier scatterometry, it should be noted that they do not show the distribution of scattered light on a hemisphere projected onto the xy-plane (as in \cite{menzel2020-BOEx}, Fig.\ 9a), but the distribution of scattered light on a plane (gnomonic projection). Therefore, the distance between the rings denoting steps of $\Delta\theta=10^{\circ}$ increases with increasing $\theta$ (cf.\ Fig.\ \ref{fig:chiasm}C).

As every image pixel represents a different position in the sample, and hence a different position with respect to the center of the LED display, the center of the scattering patterns (i.\,e.\ the region of maximum brightness where unscattered light falls straight into the camera) varies between image pixels.
In order to evaluate the scattering patterns independent of their center position, the region of maximum brightness (centroid of pixels with maximum intensity) was determined for each scattering pattern. (As the scattering patterns are not always radially symmetric, we cannot simply use the centroid of the scattering patterns.) Subsequently, the scattering patterns were cropped to the maximum possible circle around the center (cf.\ Fig.\ \ref{fig:setup}B, dashed circle). 

To quantify the distribution of scattered light and compare the resulting scattering patterns to previous results from angular SLI measurements \cite{menzel2021}, \textit{SLI profiles} (polar integrals $I(\phi)$, as in \cite{menzel2020-BOEx}) were computed: For this purpose, the intensity values were integrated from the center to the outer circle of the (centered) scattering pattern in one pixel steps for a defined azimuthal angle $\phi$, starting on top and moving clock-wise in defined steps $\Delta\phi$, using bilinear interpolation to compute the intensity value at the respective position. Figure \ref{fig:setup}B shows an example of such an SLI profile (integrated intensity values $I(\phi)$ plotted against $\phi$, for $\Delta\phi = 1^{\circ}$).
Note that, when comparing the SLI profiles to line profiles obtained from coherent Fourier scatterometry, it is only possible to compare the general form of the line profiles and the number/location of the peaks due to differences in the measurement techniques.
\\


\subsection{Smoothing of SLI Profiles}
\label{sec:methods-smoothing}

While the line profiles obtained from previous angular SLI measurements (with $\Delta\phi=15^{\circ}$ steps) are highly discretized, the line profiles obtained from coherent Fourier and SLI scatterometry measurements allow for much smaller azimuthal steps (e.\,g.\ $\Delta\phi = 1^{\circ}$) when using interpolation.
To make these line profiles analyzable with the software SLIX (originally designed to deal with highly discretized SLI profiles, see \cite{slix}), smoothing was applied to suppress high frequency components that represent details of the underlying fibers which are not relevant when characterizing the overall fiber structure \cite{menzel2020-BOEx}.

The SLI line profiles were smoothed using a Fourier low pass filter with a \textit{cutoff frequency} (determining the percentage of passing frequencies) and a \textit{window width} (controlling the sharpness of the low pass filter):
\begin{equation}
\text{FFT} \left\lbrace 1 - \left[0.5 + 0.5 \cdot \tanh \left( \frac{ \vert \text{frequencies / max. frequency} \vert - \text{cutoff frequency}}{ \text{window width}} \right) \right] \right\rbrace.
\end{equation}

To identify the optimum parameters for the filter, we considered brain regions for which it is anatomically plausible or known by design that a significant number of image pixels belongs to a certain fiber constellation (parallel in-plane fibers, out-of-plane fibers, two and three in-plane crossing fiber bundles). Note that the selected regions are not completely composed of one specific fiber constellation; they were chosen to be large enough to obtain sufficient statistics and make the optimization procedure more robust.
The line profiles obtained from SLI scatterometry measurements of these regions were filtered using different cutoff frequencies and window widths, and the \textit{detection rate}, i.\,e.\ the fraction of lines profiles for which the number of significant peaks corresponds to the respective fiber constellation (two distinct peaks for parallel fibers, four distinct peaks for two crossing fiber bundles, etc.) was determined for each pair of parameters. To avoid false positives, it was ensured that peak shape and distance also match the expectation.
For each selected region, 1600 line profiles from SLI scatterometry measurements were generated and evaluated as a black box.

\begin{figure}[h!]
\begin{center}
\includegraphics[width=0.7\textwidth]{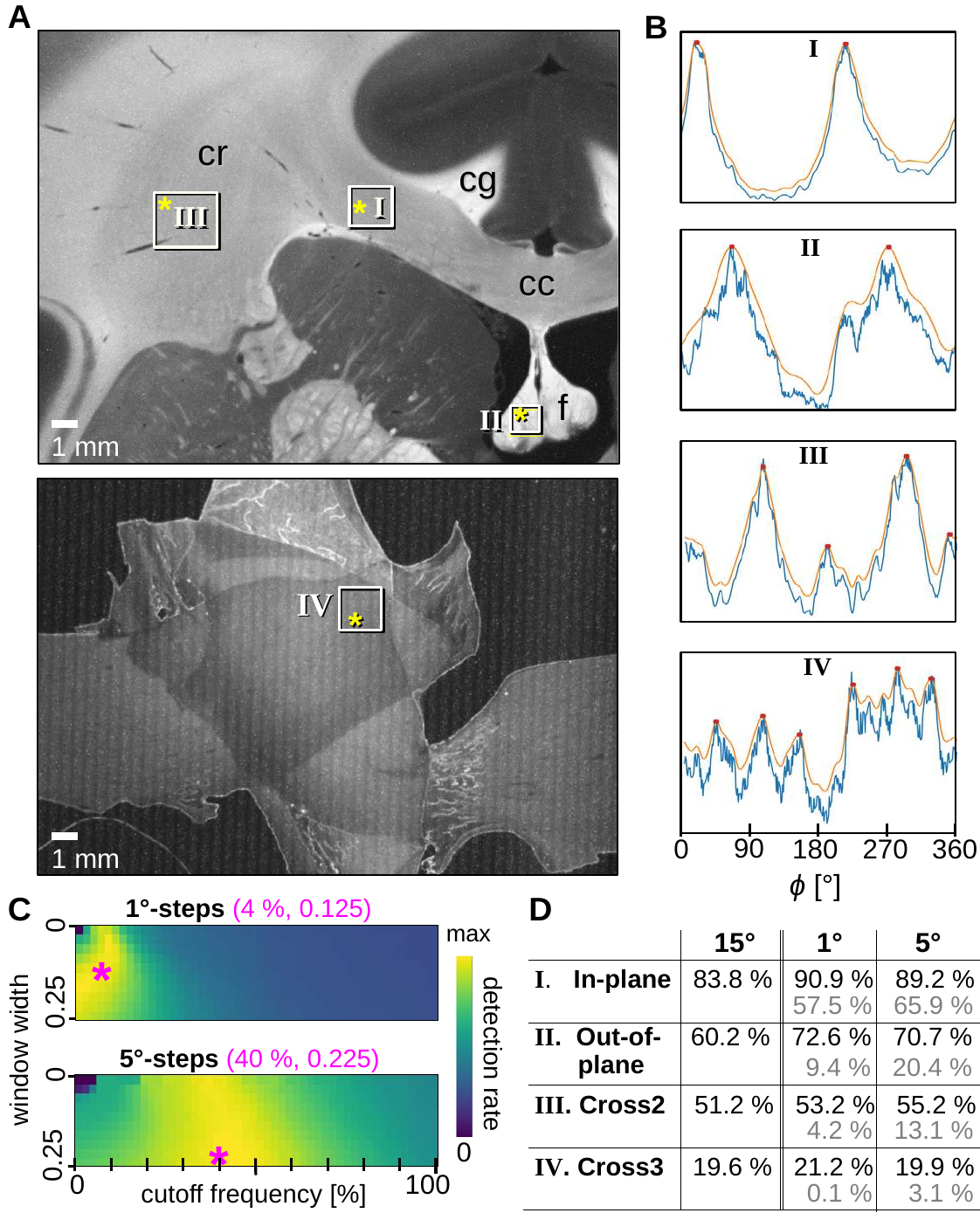}
\end{center}
\caption{Optimized smoothing of SLI profiles. \textbf{(A)} Averaged scattered light intensity of a coronal vervet monkey brain section (section 493, top) and three crossing sections of human optic tracts (sections 32/33, bottom). The white rectangles mark the evaluated regions containing mostly (I) in-plane, (II) out-of-plane, (III) two times crossing, (IV) three times crossing nerve fibers. Relevant anatomical structures are labeled: corpus callosum (cc), cingulum (cg), corona radiata (cr), fornix (f). \textbf{(B)} Examples of original (blue) and smoothed (orange) SLI profiles (normalized by their maximum value) with $\Delta\phi = 1^{\circ}$-steps, obtained from scattering patterns measured at locations indicated by the yellow asterisks in A. The SLI scatterometry measurements were performed with one illuminated LED and 10 sec illumination. For the vervet brain section, the measurement was performed 16 months after tissue embedding with $64 \times 64$ kernels and a gain factor of 27. For the three sections of optic tracts, the measurement was performed 20 months after tissue embedding with $50 \times 50$ kernels and a gain factor of 10. \textbf{(C)} Detection rate (average over the four selected tissue types) for different parameters of the Fourier low pass filter (different cutoff frequencies and windows widths) applied to the SLI profiles generated with $\Delta\phi=1^{\circ}$-steps (top) and $5^{\circ}$-steps (bottom). The magenta asterisks mark the set of parameters (shown in magenta numbers) for which the maximum detection rate is reached: 83.4\,\% for $1^{\circ}$-steps and 80.8\,\% for $5^{\circ}$-steps. \textbf{(D)} Detection rates evaluated separately for the different regions in A. The black numbers show the detection rates for SLI profiles with $15^{\circ}$-steps (without smoothing) as well as for SLI profiles with $1^{\circ}$- and $5^{\circ}$-steps when using the optimum smoothing parameters (magenta numbers in C). The gray numbers show the detection rates before applying the smoothing.}
\label{fig:optimization}
\end{figure}

Figure \ref{fig:optimization} shows the optimization of the filter parameters. 
For azimuthal steps of $\Delta\phi = 1^{\circ}$ and $5^{\circ}$, the Fourier low pass filter was applied to all line profiles and optimized by shifting both the cutoff frequency and the window width of the filter for each iteration. 
For the cutoff frequency, steps of $2\,\%$ were selected. 
The window width was incremented in steps of $0.025$ for a range between $0.00$ and $0.25$. 
The exact implementation of the Fourier low pass filter is described in the software repository of SLIX, which is publicly available (\url{https://github.com/3d-pli/SLIX}).

For each region, the optimization algorithm yields a matrix with computed detection rates.  
Each fiber arrangement has its own optimum set of parameters: For example, two or three crossing fibers (expected to generate line profiles with four or six peaks\footnote{Note that the optimization procedure only considers the total number of determined peaks. If one peak is not detected (e.\,g.\ five instead of six peaks), the other (e.\,g.\ two crossing) orientations could still be correctly determined although the line profile does not count in the detection rate.}) need a higher cutoff frequency, i.\,e.\ higher passing frequencies, than parallel in/out-of-plane fibers (expected to generate one or two peaks). 
Also, crossing regions, which were less correctly determined in previous angular SLI measurements \cite{menzel2021}, may yield lower values in the corresponding matrix than their in/out-of-plane counterparts. 
To ensure that each type of fiber region is equally considered in the selection of optimum filter parameters, all matrices were normalized and summed up to identify the best choice of parameters.

The results are shown in Fig.\ \ref{fig:optimization}C for both azimuthal steps. 
Regions in yellow have higher detection rates than blue areas. 
The parameter combination with the highest value, indicated by the magenta asterisk, was chosen as optimum filter parameters for the given azimuthal step: Line profiles with $\Delta\phi = 1^{\circ}$ were filtered with a cutoff frequency of $4\,\%$ and a window width of 0.125 (used in Figs.\ \ref{fig:optimization}B, \ref{fig:chiasm}D, \ref{fig:vervet-SLISc}D). Line profiles with $\Delta\phi = 5^{\circ}$ were filtered with a cutoff frequency of $40\,\%$ and a window width of 0.225 (used in Fig.\ \ref{fig:human}B).  
The effect of the Fourier low pass filter can be seen in Fig.\ \ref{fig:optimization}D where the detection rate\footnote{Note that the given numbers are the result of the optimization algorithm and are only indirectly related to the accuracy of reconstruction. As the algorithm optimized the detection rate for each selected brain region separately, the numbers should only be compared line by line.} of unfiltered $15^{\circ}$-line profiles (used in Figs.\ \ref{fig:chiasm}A/B and \ref{fig:vervet-SLI}B,C) is compared to the detection rate of filtered $1^{\circ}$- and $5^{\circ}$-line profiles (in black): The filtered line profiles obtained from SLI measurements with small azimuthal steps yield comparable or better detection rates than those obtained from the highly discretized SLI measurements. The gray numbers show the detection rates when no filter is applied, demonstrating that the smoothing of the $1^{\circ}$- and $5^{\circ}$-line profiles greatly increases the detection rate. \\


\subsection{Visualization of Nerve Fiber Directions}
\label{sec:methods-visualization}

The smoothed SLI profiles were analyzed with SLIX \cite{slix} to determine the positions of the peaks and compute the corresponding nerve fiber orientations as described by \cite{menzel2021}: The in-plane fiber orientations were computed from peak pairs with a distance of $(180 \pm 35)^{\circ}$.
The previous angular SLI measurements shown in this paper were evaluated with SLIX as described in \cite{menzel2021}, using no smoothing and the centroid of the peak tips in the line profiles to improve upon the angular resolution of $15^{\circ}$.

\begin{figure}[h]
\begin{center}
\includegraphics[width=\textwidth]{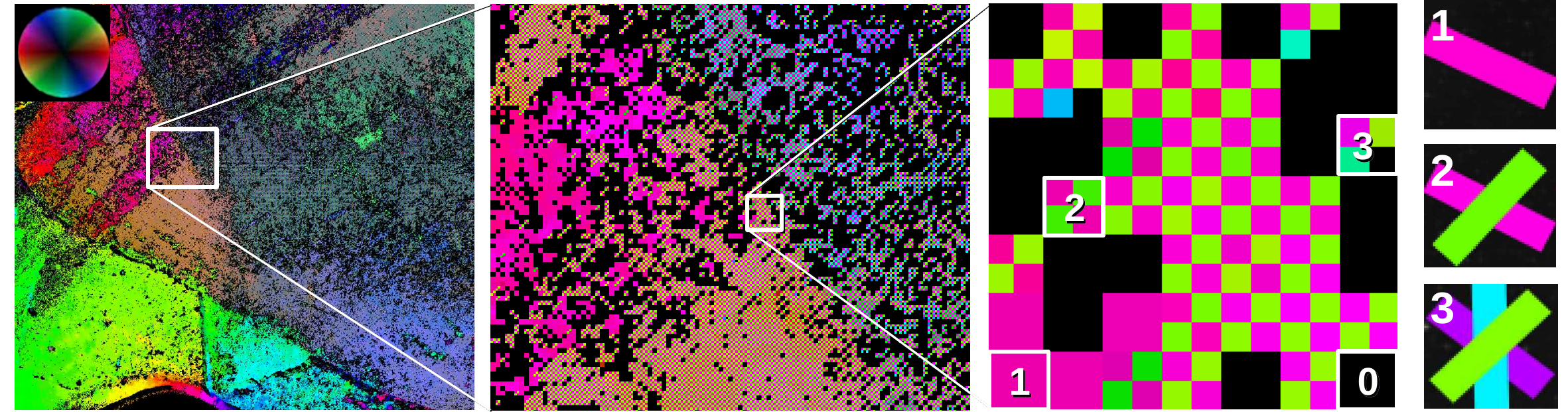}
\end{center}
\caption{Visualization of fiber orientations, shown exemplary for a region in Fig.\ \ref{fig:chiasm}A: The in-plane fiber orientations are displayed by different colors (color wheel at the top left). Every image pixel contains $2 \times 2$ subpixels with up to four different colors, depending on the number of derived fiber orientations. Here, up to three different fiber orientations are shown (see enlarged region). Image pixels for which no fiber direction could be determined are displayed in black. Individual fiber orientations can be represented as colored lines (as shown on the right).} 
\label{fig:visualization}
\end{figure}

The \textit{fiber orientation map} is a simple way to visualize and interpret SLI measurements (see Fig.\ \ref{fig:visualization}): Each in-plane fiber orientation (direction angle) is mapped to a unique color based on the hue channel of the HSV color space (see color wheel at the top left). For each measured image pixel, up to four different fiber directions are stored in $2 \times 2$ subpixels. Depending on the number of derived fiber orientations, the subpixels have one, two, three, or four different colors (see enlarged region on the right). 
Image pixels with no determined fiber direction are displayed in black.
With this approach, crossing regions are immediately visible in the fiber orientation map and no information is lost.

Another kind of visualization is the \textit{fiber orientation distribution map} as shown in Figs.\ \ref{fig:chiasm}B, \ref{fig:vervet-SLI}C, and \ref{fig:vervet-SLISc}D. 
This visualization sacrifices single-pixel accuracy in order to display the individual fiber orientations as unit vectors.
For this purpose, the determined fiber direction angles are converted to unit vectors for each image pixel and plotted as colored lines (cf.\ Fig.\ \ref{fig:visualization} on the very right).
To make the fiber orientations visible for a large area, the vectors are displayed for less image pixels.
However, without reducing the number of displayed vectors, the result will be somewhat equal to the fiber orientation map. 
Our approach is to increase the size of the vectors and keep the information density high at the same time. 
Hence, instead of thinning out the vector image and showing the vectors e.\,g.\ for every $40^{\text{th}}$ image pixel, all unit vectors in the region (e.\,g.\ the vectors of $40 \times 40$ image pixels) are shown on top of each other. 
To identify the dominating fiber orientation for each region, the vectors are assigned a low alpha value. 
Thus, a single vector with a different orientation than the dominating orientation appears very faintly, but is still visible.

\section{Results}

\subsection{Calibration Measurements}
\label{sec:comparison}

\begin{figure}[h!]
\begin{center}
\includegraphics[width=\textwidth]{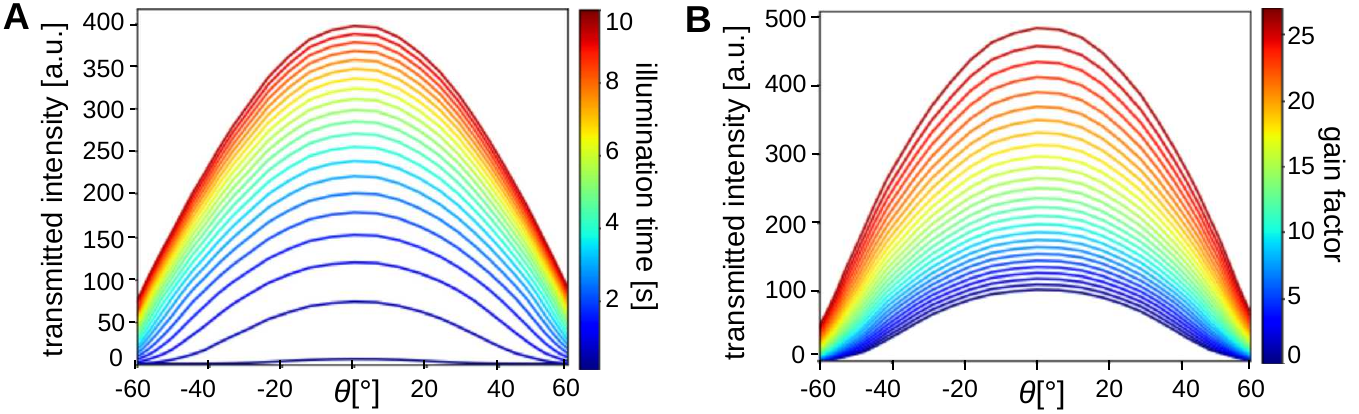}
\end{center}
\caption{Average transmitted light intensity of a diffusor plate (inner $1000 \times 1000$ px) illuminated by $8 \times 8$ LEDs from 32 different positions (along the center line of the LED display; $\theta=0^{\circ}$ corresponds to the middle position of the display, cf.\ Fig.\ \ref{fig:setup}A). The different curves belong to measurements with (\textbf{A}) different illumination times (100\,ms, 500\,ms, 1000\,ms, 1500\,ms, $\dots$, 10\,s) with gain 3, and (\textbf{B}) different gain factors (1, 2, ..., 27) with 1\,sec illumination.} 
\label{fig:calibration}
\end{figure}

To study the radiation characteristics of the different LEDs and estimate their impact on the measured scattering patterns, calibration measurements were performed. For this purpose, a diffusor plate (with homogeneous scattering properties) was placed in the sample holder and illuminated by a square of $8 \times 8$ LEDs. The square was moved along the center line of the LED display and an image was recorded for each of the 32 different positions ($x = -15.5, ..., 15.5$). The average light intensity of the inner $1000 \times 1000$ image pixels was plotted against the angle of illumination: $\theta = \arctan\left(\frac{x\,\cdot\,8\,\cdot\,1.8\,\text{mm}}{13\,\text{cm}}\right)$.
Figure \ref{fig:calibration} shows the corresponding curves for different illumination times (A) and different gain factors of the camera (B).

When illuminating with larger angles, the transmitted light intensity decreases significantly, especially for large illumination times and gain factors. This can be explained by the fact that the distance between light source and sample increases with increasing illumination angle (the intensity decreases $\propto 1/r^2$ for spherical emitters) and the LEDs have a limited angle of radiation (view angle $\lesssim 120^{\circ}$) so that outer LEDs do not emit much light under large angles. An illumination time of more than one second is needed to achieve sufficient transmitted light intensities when illuminating from the outer border of the display.
Note that the diffusor plate leads to more scattering and absorption than an object carrier with brain section. The required illumination time is therefore expected to be shorter (or comparable when using a smaller number of illuminated LEDs).
The illumination characteristics should be taken into account when interpreting the measured scattering patterns. For the resulting line profiles, which are computed by integrating from the center to the outer border of the scattering pattern, we do not expect qualitative changes (concerning e.\,g.\ the peak positions).

\subsection{Comparison to Previous SLI Measurements, Coherent Fourier Scatterometry, and Simulations}
\label{sec:comparison}

\begin{figure}[h!]
\begin{center}
\includegraphics[width=\textwidth]{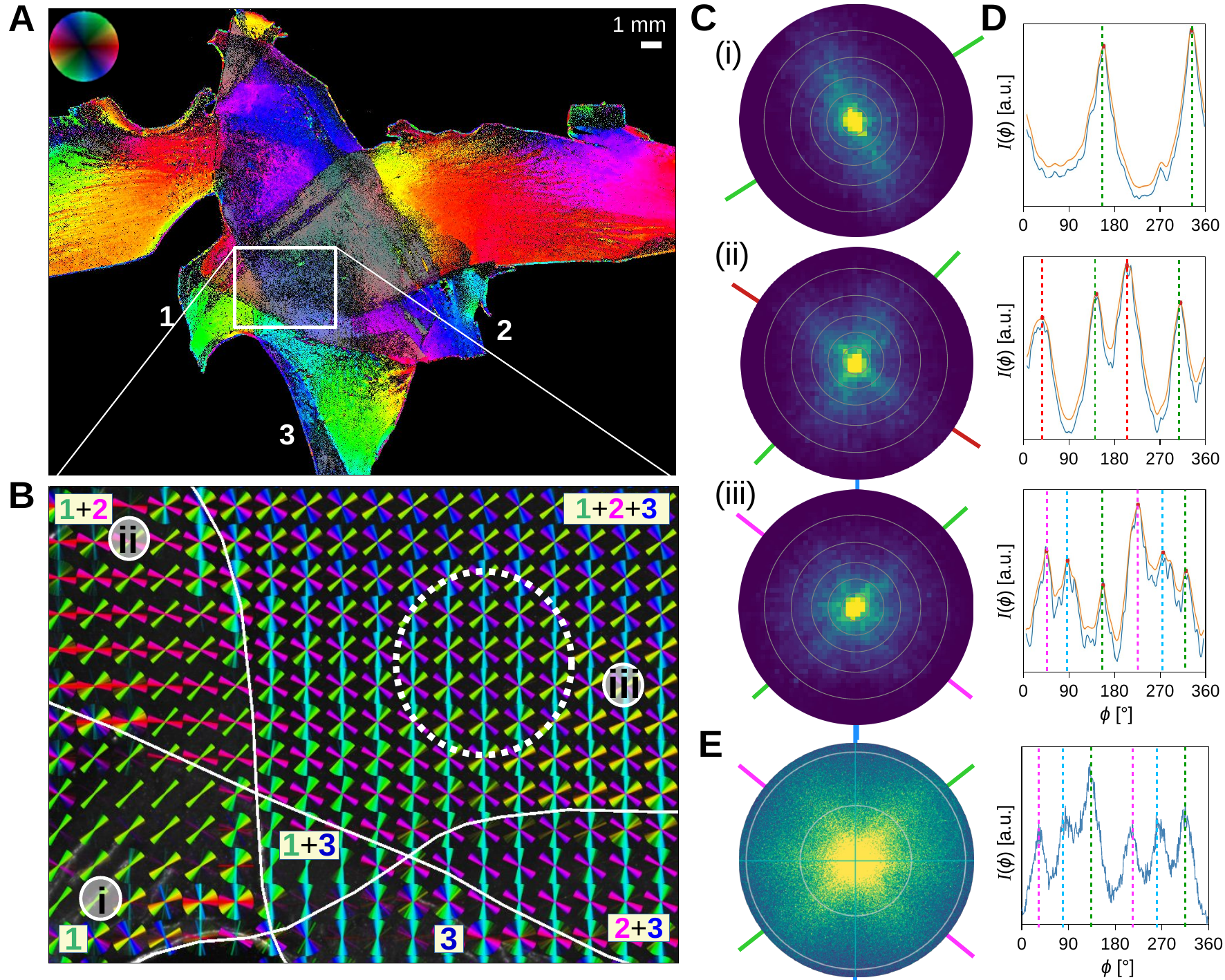}
\end{center}
\caption{Three crossing sections of human optic tracts measured with angular/scatterometry SLI. \textbf{(A,B)} In-plane fiber orientations obtained from an angular SLI measurement with $15^{\circ}$ azimuthal steps, 1 sec illumination, and px = 6.5\,\textmu m, performed 4 months after tissue embedding (as shown in \cite{menzel2021} Fig.\ 5, bottom). The fiber orientations are shown for each image pixel in different colors (A), and for $40 \times 40$ image pixels together as colored lines (B). The three tissue layers are marked by arabic numbers. \textbf{(C,D)} Scattering patterns and corresponding (smoothed) azimuthal line profiles obtained from an SLI scatterometry measurement of the same sample ($2\times2$ illuminated LEDs, $50 \times 50$ kernels, gain factor 10, illumination 5 sec, 15 months after tissue embedding), evaluated exemplary for image pixels located in one (i), two (ii), and three (iii) tissue layers, at the positions indicated in B. The concentric rings in the scattering patterns denote steps of $\Delta\theta = 10^{\circ}$. The dashed vertical lines in D indicate the determined peak positions from which the nerve fiber orientations were computed (indicated by solid lines in C in the respective color). \textbf{(E)} Scattering pattern and corresponding azimuthal line profile obtained from a coherent Fourier scatterometry measurement for the region marked by the dashed white circle in B (adapted from \cite{menzel2020-BOEx}, Fig.\ 5(iv); the measurement was performed with a laser with 1.12\,mm diameter, a numerical aperture of 0.4, and 4 months after tissue embedding).}
\label{fig:chiasm}
\end{figure}

\begin{figure}[h!]
\begin{center}
\includegraphics[width=\textwidth]{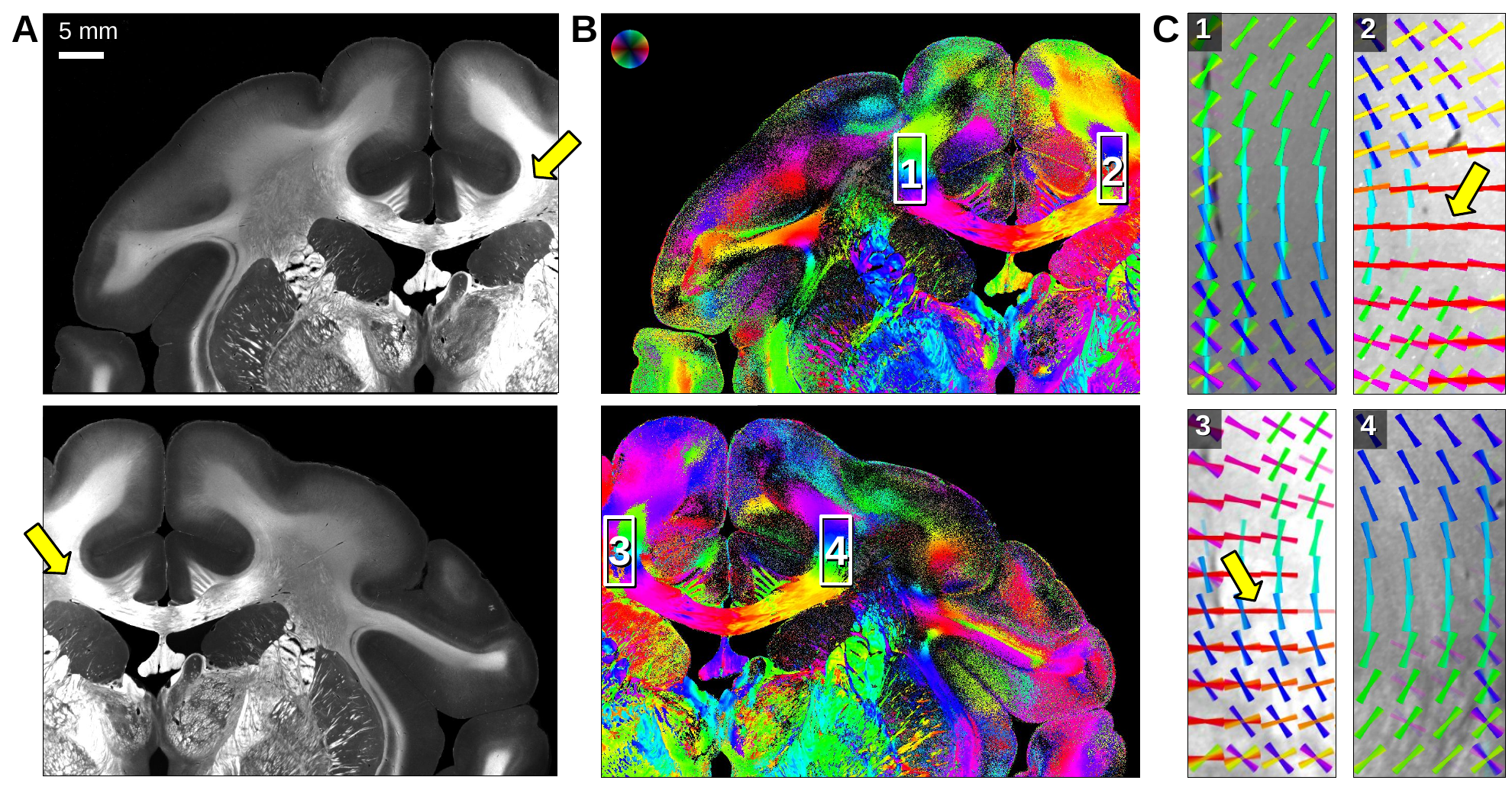}
\end{center}
\caption{Angular SLI measurements of a coronal vervet monkey brain section (section 512) for different fields of view (left/right hemisphere). The measurements were performed one day after tissue embedding with $\Delta\phi=15^{\circ}$-steps, 0.5 sec illumination, and px = 13.7\,\textmu m. \textbf{(A)} Maximum scattered light intensity. \textbf{(B)}
In-plane fiber orientations displayed for each image pixel in different colors. \textbf{(C)} Fiber orientation distribution maps of the regions highlighted in B: fiber orientations are displayed on top of each other as colored lines for every $30 \times 30$ image pixels. The arrows mark artifacts caused by asymmetric illumination of the respective regions.}
\label{fig:vervet-SLI}
\end{figure}

\begin{figure}[h!]
\begin{center}
\includegraphics[width=0.9\textwidth]{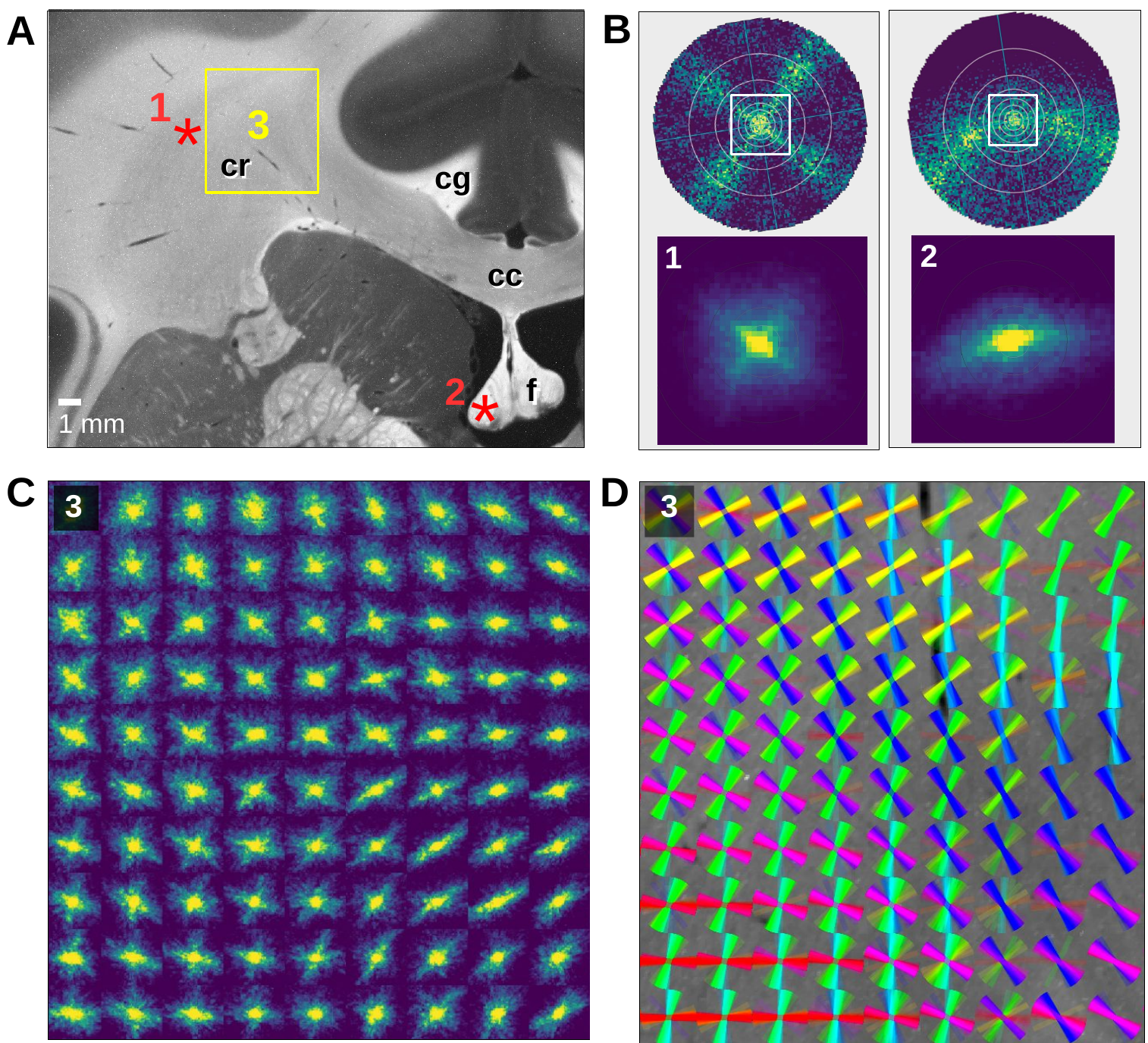}
\end{center}
\caption{SLI scatterometry measurements of a coronal vervet monkey brain section (section 493). \textbf{(A)} Averaged scattered light intensity with labeled anatomical structures: corpus callosum (cc), cingulum (cg), corona radiata (cr), fornix (f). \textbf{(B)} Scattering patterns for two crossing fiber bundles (left) and an out-of-plane fiber bundle (right). The top images show the simulated scattering patterns obtained from finite-difference time-domain simulations of two $90^{\circ}$-crossing, interwoven fiber bundles and a $50^{\circ}$-inclined fiber bundle (adapted from \cite{menzel2020}, Fig.\ 7). The bottom images show the measured scattering patterns for an image pixel in the corona radiata (1) and in the fornix (2), indicated by the red asterisks in A. The SLI scatterometry measurement was performed 10 months after tissue embedding with $4 \times 4$ illuminated LEDs, $40 \times 40$ kernels, gain factor 10, and illumination 10 sec.
\textbf{(C)} Scattering patterns of the rectangular region in A, shown for every $150^{\text{th}}$ image pixel (px = 3\,\textmu m). The SLI scatterometry measurement was performed 15 months after tissue embedding with one illuminated LED, $50 \times 50$ kernels, gain factor 27, and illumination 10 sec. \textbf{(D)} Fiber orientation distribution map of the same region: the fiber orientations were computed with SLIX from every $15^{\text{th}}$ scattering pattern and displayed on top of each other as colored lines for every $10 \times 10$ scattering patterns.}
\label{fig:vervet-SLISc}
\end{figure}

To validate the method of SLI scatterometry and enable a direct comparison to previous results from angular SLI measurements and coherent Fourier scatterometry, the same samples were used for the measurement: three crossing sections of human optic tracts (Fig.\ \ref{fig:chiasm}) and coronal vervet monkey brain sections (Figs.\ \ref{fig:vervet-SLI} and \ref{fig:vervet-SLISc}). 

Figure \ref{fig:chiasm}C,D shows the SLI scattering patterns and corresponding line profiles for three selected image pixels in one, two, and three crossing tissue layers of the optic tracts. The scattering patterns show the scattering behavior of in-plane crossing nerve ﬁbers as predicted by simulations \cite{menzel2020} and observed in coherent Fourier scatterometry \cite{menzel2020-BOEx}: In-plane nerve ﬁbers generate scattering reflexes perpendicular to their orientation so that the orientations of crossing nerve ﬁbers can be determined by the position of the peaks in the resulting line profiles. As expected, the scattering patterns show one, two, and three scattering reflexes which correspond to two, four, and six distinct peaks in the line profiles. From the positions of the peaks, the orientations of the nerve fibers were computed as described in Sec.\ \ref{sec:methods-visualization} and visualized as colored lines. The fiber orientations correspond very well to the fiber orientations computed from previous angular SLI measurements at the approximate same locations (see (i),(ii),(iii) in Fig.\ \ref{fig:chiasm}). 

A further comparison to a scattering pattern obtained from coherent Fourier scatterometry of the same location (triple tissue layers, dashed circle in B) shows that the resulting line profiles are very similar to each other (cf.\ Fig.\ \ref{fig:chiasm}C(iii) and E). 
It should be noted that SLI scatterometry shows the distribution of scattered light onto a plane (leading to much lower intensities at the borders of the scattering pattern), while coherent Fourier scatterometry shows the distribution of scattered light onto a hemisphere (projected onto the xy-plane). Also, the maximum angle of illumination is different (steps of $\Delta\theta=10^{\circ}$ are marked by concentric rings in the scattering patterns, respectively). 
Taking into account that the measurements were performed at different times (A/B: three months, C-E: 10-15 months after tissue embedding), the results correspond very well to each other, demonstrating that SLI scatterometry yields reliable scattering patterns.

A major drawback of angular SLI measurements is that pixels at the outer border of the image are illuminated under different polar angles $\theta$ than pixels in the center of the image. This asymmetric illumination at the image borders leads to asymmetries in the resulting SLI profiles so that peaks might not be detected and wrong/perpendicular fiber orientations are computed. Figure \ref{fig:vervet-SLI} shows angular SLI measurements of a coronal vervet monkey brain section (section 512, as shown in \cite{menzel2021} Fig.\ 8c) for different fields of view.
The asymmetric illumination at the image borders becomes clearly visible in the maximum scattered light intensities (arrows in Fig.\ \ref{fig:vervet-SLI}A). The vector distribution maps in C show mirror-inverted regions at the borders of the corpus callosum:
In regions that are not located at the image border (1,4), the reconstructed fiber orientations follow the course of the bundle as expected. At the image borders (2,3), some reconstructed fiber orientations run perpendicularly to the course of the bundle (highlighted by arrows). 
While angular SLI yields only line profiles, SLI scatterometry provides the full structural information of the scattering patterns, allowing to reliably determine the center of illumination for each image pixel (cf.\ Fig.\ \ref{fig:setup}B) and to avoid artifacts caused by asymmetric illumination.

Figure \ref{fig:vervet-SLISc} shows SLI scatterometry measurements of a neighboring coronal vervet brain section. 
The complex nerve fiber crossings in the corona radiata (cr) are nicely visible, both in the scattering pattern map (C) and in the corresponding fiber orientation distribution map (D). Figure \ref{fig:vervet-SLISc}B(1) shows the SLI scattering pattern of one image pixel in the corona radiata (as indicated in A) in direct comparison to a simulated scattering pattern for two $90^{\circ}$-crossing fiber bundles (\cite{menzel2020}, Fig.\ 7b). Taking the decrease of the scattered light intensity with increasing illumination angle (cf.\ Fig.\ \ref{fig:calibration}) and the smaller angular range (indicated by the white square) into account, the measured scattering pattern corresponds very well to the simulated one.

In addition to fiber crossings, SLI scatterometry allows to study out-of-plane fiber structures.
Figure \ref{fig:vervet-SLISc}B(2) shows the SLI scattering pattern obtained from an image pixel in the fornix (as indicated in A), where most fibers are running out of the coronal section plane. The corresponding top figure in B shows the simulated scattering pattern of a fiber bundle with an out-of-plane inclination angle of $50^{\circ}$. While in-plane nerve fibers generate scattering reflexes perpendicular to their orientation, the simulation studies \cite{menzel2020} predict that the light is scattered more and more in the direction of the fibers with increasing out-of-plane angle. As expected, the SLI scattering pattern shows a slightly curved reflex, demonstrating again that SLI scatterometry does not only yield compatible results with previous measurements (angular SLI and coherent Fourier scatterometry), but also with simulation results and theoretical predictions.
\\

\subsection{Evaluation of Human Brain Section}
\label{sec:human-section}

\begin{figure}[h!]
\begin{center}
\includegraphics[width=0.8\textwidth]{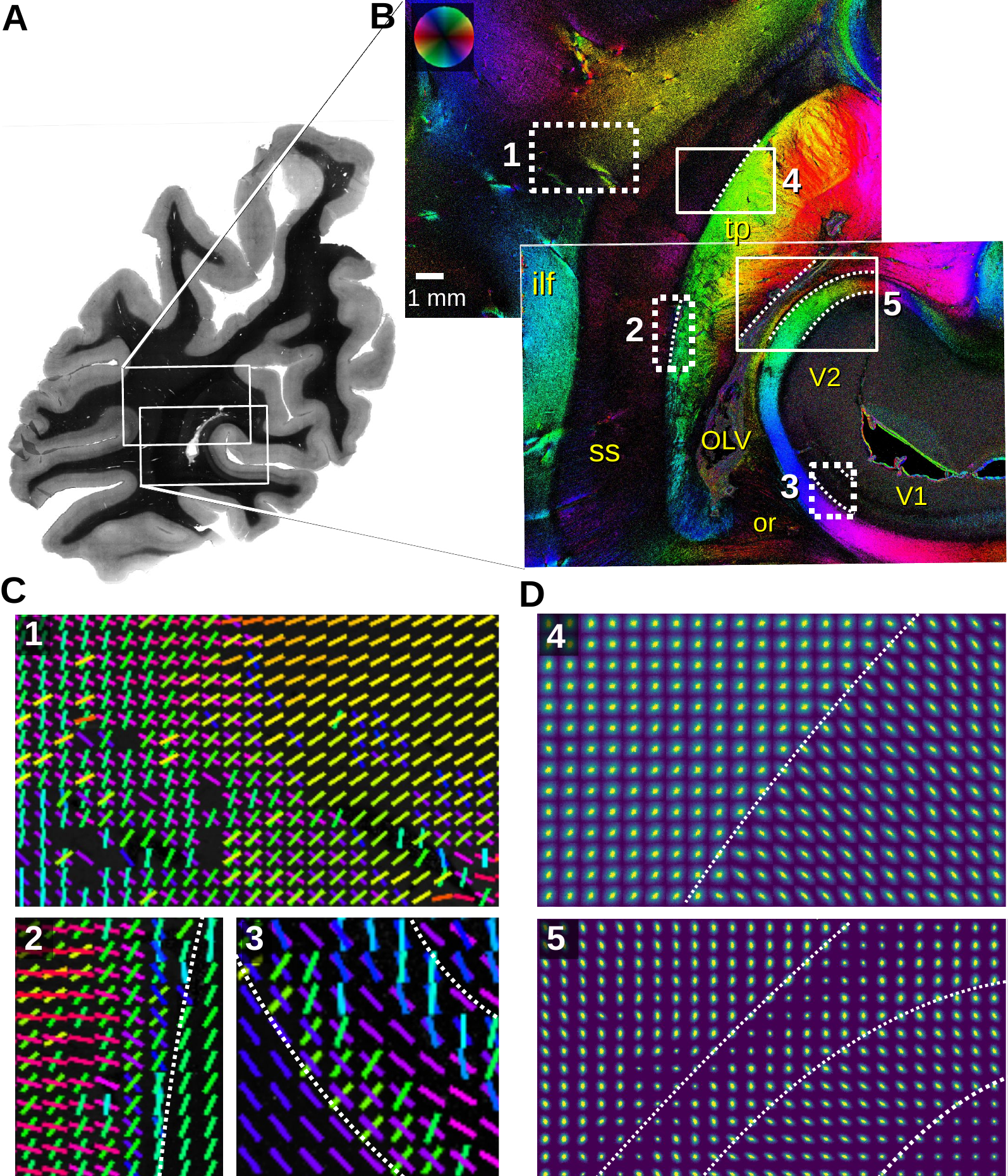}
\end{center}
\caption{Section of a left human hemisphere measured with angular/scatterometry SLI. \textbf{(A)} Transmittance image. \textbf{(B)} In-plane fiber orientations obtained from an angular SLI measurement with $\Delta\phi=5^{\circ}$ steps/width, $\theta = [42.5,47.5]^{\circ}$, px = 3\,\textmu m, performed 8 months after tissue embedding with green light (gain factor 2, illumination 3 sec). Relevant anatomical structures are labeled in yellow (ilf: inferior longitudinal fascicle, OLV: occipital lateral ventricle, or:  optic radiation, ss:  sagittal stratum, tp: tapetum, V1: primary visual cortex, V2: prestriate visual cortex). \textbf{(C)} Fiber orientation vectors shown for every $40^{\text{th}}$ image pixel for the regions (1,2,3) marked in B. Vectors are shown if at least 8\,\% of the surrounding $40 \times 40$ pixels have a defined fiber orientation. \textbf{(D)} Scattering pattern maps obtained from SLI scatterometry measurements of the same sample ($4\times4$ illuminated LEDs, $64 \times 64$ kernels, gain factor 10, illumination 3 sec, 8 months after tissue embedding), shown for every $50^{\text{th}}$ image pixel for the regions (4,5) marked in B. The white dashed lines indicate tissue borders for better reference.}
\label{fig:human}
\end{figure}

Figure \ref{fig:human} shows the results of an angular SLI measurement with azimuthal steps of $5^{\circ}$ (B,C) and an SLI scatterometry measurement (D) for a coronal section of a left human brain hemisphere (A). 

The fiber orientation maps in B show the \textit{sagittal stratum (ss)} and surrounding white matter structures. The sagittal stratum contains highly parallel nerve fibers running mostly in rostro-caudal direction, perpendicular to the coronal section plane. Its environment, however, is characterized by fiber bundles, e.\,g.\ \textit{inferior longitudinal fascicle (ilf)} and \textit{tapetum (tp)}, entering or perforating the sagittal stratum from its lateral and medial interface.
Multiple fiber directions in the white dashed rectangles are displayed as colored lines for every $40^{\text{th}}$ image pixel in C. The fiber orientation vectors reveal astonishing details, even of very small fiber bundles: The fibers of the inferior longitudinal fascicle (cyan) in region (1), for instance, join fibers originating from the dorso-parietal cortex (red fibers) and proceed dorso-medially to the parasagittal cortex (yellow). However, another portion of the red fibers branches off, crossing the inferior longitudinal fascicle, and enters the sagittal stratum (magenta).
Region (2) reveals fiber bundles splitting off the tapetum on their ventral course to the lateral side (green, yellow), while other fibers coming from the lateral side and perforating the whole sagittal stratum turn ventrally, eventually entering the tapetum (red, magenta).
The vectors in region (3) retrace the typical fiber architecture of the \textit{primary visual cortex (V1)} with the radial input (green, cyan) and the pronounced transversal fiber layer (blue, magenta) located in the center of the cortex (\textit{Stria of Gennari}).
 
The scattering pattern maps in D show the transition zone between in-plane (tapetum) and out-of-plane fiber bundles (sagittal stratum) (4) as well as for different tissue types (5), as marked in B. The scattering patterns are shown for every $50^{\text{th}}$ image pixel; the white dashed lines mark tissue borders for better reference. The transition becomes nicely visible in the different shapes of the scattering patterns: in-plane fibers show elongated scattering reflexes perpendicular to their orientation (lower right corner in 4, right stripe in 5), while out-of-plane fibers show broader, almost circular scattering reflexes (upper left corner in 4, middle stripe in 5). 

In box (5), four areas can be distinguished by their scattering pattern size and anisotropy: Very weak scattering in the cortex (V2 in the lower right corner) does not allow for the detection of distinct scattering pattern anisotropy due to the low amount of myelinated fibers. In the white matter substrate, however, the radially oriented anisotropy increases considerably dominated by the tangential course of strongly myelinated terminals of the optic radiation. The next layer consists of terminals of the major forceps corporis callosi (medial layer), constituting the medial wall of the occipital horn of the lateral ventricle (lateral layer). The major forceps is characterized by a vertical scattering anisotropy. The lateral ventricle can be identified by a single strand of tiny isotropic scattering patterns. Eventually, the tapetum is again characterized by strong scattering with vertically oriented anistropy identifying transversal fibers (red area in the upper left area of 5).

The results demonstrate that SLI scatterometry is a powerful approach, revealing the intricate nerve fiber architecture of the human brain.\\


\section{Discussion}

The presented scatterometry measurements with Scattered Light Imaging (SLI) allow the simultaneous generation of complete scattering patterns for all image pixels in an investigated brain section. In this way, SLI scatterometry provides full structural information of complex nerve fiber structures, even in brain regions with densely packed fibers.

Coherent Fourier scatterometry \cite{menzel2020-BOEx} measures the complete scattering patterns with high detail, but requires mechanical rasterizing of the sample and the object-space resolution is limited to the minimum diameter ($> 100$\,\textmu m) of the laser beam. With SLI scatterometry, a scattering pattern can be measured for every image pixel so that the object-space resolution is only limited by the available optics, which could still be improved (here: px = 3\,\textmu m). The resolution of the scattering patterns, determining the accuracy with which the underlying nerve fiber structures can be reconstructed, is determined by the density of LEDs (here:1.8\,mm pixel pitch) and the number of measurements (e.\,g.\ $40 \times 40$ kernels). 

We could show that the SLI scattering patterns are scatterometry and agree with predictions from simulation studies (see Sec.\ \ref{sec:comparison}). So far, this correspondence was only shown for line profiles \cite{menzel2021}, not for whole scattering patterns.
This demonstrates that SLI scatterometry can be used to measure scattering patterns for each image pixel at once, and to make valid predictions for the underlying nerve fiber structures.

Previous angular SLI measurements using 24 azimuthal illumination angles with $\Delta\phi = 15^{\circ}$ steps and fixed polar angle \cite{menzel2021} yield highly discretized line profiles and provide not enough details for a comprehensive analysis of the underlying fiber structures. 
The scattering patterns obtained from SLI scatterometry enable a much more reliable and robust interpretation of the scattering signal because more data points (image pixels) are included: Using integration and bilinear interpolation, highly resolved SLI profiles can be generated (down to $\Delta\phi=1^{\circ}$, see Sec.\ \ref{sec:methods-scatteringpatterns}), allowing for a more accurate determination of the in-plane nerve fiber orientations. 
A significant improvement of SLI scatterometry over angular SLI measurements is that it allows to identify the center of the scattering pattern before computing the SLI profiles, thus avoiding artifacts from asymmetric illumination at the image borders (cf.\ arrows in Fig.\ \ref{fig:vervet-SLI}C).

The current evaluation of the scattering patterns is limited to the generation of line profiles (fiber orientation distribution maps), which are by design similar to those obtained from angular SLI measurements. In contrast to the line profiles, the scattering patterns provide the full structural information of the brain tissue, allowing for more advanced studies e.\,g.\ in case of out-of-plane nerve fibers: The scattering patterns show a bending of the scattering reflex (cf.\ Fig.\ \ref{fig:vervet-SLISc}B(2)), which is easier to fit than the distance of merging peaks in an SLI profile, and will allow for a more reliable determination of the out-of-plane fiber angles in future work. Also, when determining e.\,g.\ crossing inclined nerve fibers (cf.\ Fig.\ F3 in \cite{menzel2021}), scattering patterns will be easier to interpret than SLI profiles, which can be ambiguous. Future work should also exploit other characteristics of the scattering patterns (e.\,g.\ the curvature of the scattering reflexes or the maximum scattering angles) to extract additional information like the homogeneity of the tissue.
A comparison of scattering patterns obtained from measurements with different wavelengths (red, green, blue) could reveal the size of scattering structures such as the fiber diameters.

While angular SLI measurements require no more than 72 images and take less than 4 min (for $\Delta\phi=5^{\circ}$), the here presented SLI scatterometry measurements require thousands of images and take several hours (a measurement with $64 \times 64$ kernels and 3 sec illumination, as shown in Fig.\ \ref{fig:human}D, takes at least 3.4 hours).
We decided to perform the SLI scatterometry measurements with a large number of images (illuminating with single LEDs at times) to generate scattering patterns with high structural detail which can be used to validate our technique and show that we obtain similar patterns as coherent Fourier scatterometry, but for each image pixel at once and with microscopic resolution (px = 3\,\textmu m). In future work, these high-resolution scattering patterns can be used as ground truth in order to develop more efficient measurement protocols that reduce the number of required images while maintaining accuracy and resolution of the resulting scattering patterns. Resampling the high-resolution SLI scattering patterns in different ways allows to study how much the number of images can be reduced without losing important details in the generated scattering patterns. Another idea is to use concepts from compressed sensing \cite{duarte2011} and exploit the sparsity of the signal in order to significantly reduce the amount of images to recover the signal. In SLI scatterometry, every camera pixel can be understood as an independent single pixel camera \cite{duarte2008} so that the same approaches from compressed sensing can be applied. As these approaches employ the illumination of many LEDs, this will also reduce the required illumination time.


\section*{Conflict of Interest Statement}


The authors declare that the research was conducted in the absence of any commercial or financial relationships that could be construed as a potential conflict of interest.


\section*{Author Contributions}

MM designed and supervised the research. MR conducted the calibration measurements. JR programmed the smoothing of the SLI profiles and visualized the fiber orientation and vector maps. MR and DG performed the SLI scatterometry measurements. DG assisted with anatomical labeling. MM prepared the results for the figures and wrote the first draft of the manuscript. JR and DG wrote sections of the manuscript. All authors contributed to manuscript revision, read, and approved the submitted version.


\section*{Funding}

This work was funded by the Helmholtz Association port-folio theme ``Supercomputing and Modeling for the Human Brain'', the European Union’s Horizon 2020 Research and Innovation Programme under Grant Agreement No.\ 945539 (``Human Brain Project'' SGA3), and the National Institutes of Health under grant agreements No.\ R01MH092311 and 5P40OD010965.


\section*{Acknowledgments}
The authors gratefully thank Roxana Kooijmans (Netherlands Institute for Neuroscience, Amsterdam) for providing the human brain tissue, Karl Zilles and Roger Woods (David Geffen School of Medicine at UCLA, USA) for collaboration in the vervet brain project, and the lab team of the INM-1 (Forschungszentrum J\"ulich GmbH, Germany) for preparing the brain sections.


\section*{Data Availability Statement}
The software \textit{SLIX} \cite{slix} that was used for the smoothing of the line profiles and for the visualization of the nerve fiber orientations is available on GitHub (\url{https://github.com/3d-pli/slix}).


\bibliography{BIBLIOGRAPHY}


\end{document}